\documentclass[english,preprint,prb,onecolumn,showpacs,preprintnumbers,amsmath,amssymb,showkeys]{revtex4-2}

\usepackage[utf8]{inputenc}
\usepackage[T1]{fontenc}
\usepackage{graphicx}
\usepackage{xcolor}
\usepackage{bm}
\usepackage[colorlinks, linkcolor= blue, citecolor = blue, urlcolor=blue]{hyperref}
\usepackage{physics}
\usepackage{float}
\usepackage{wrapfig}
\usepackage[stable]{footmisc}
\usepackage{array}
\usepackage{array}
\usepackage{textcomp}
\usepackage{multirow}
\usepackage{hhline}
\usepackage{pifont}
\usepackage{chemformula}
\usepackage{textcase}
\usepackage{ulem}
\usepackage{bm}
\usepackage{color}
\usepackage{amssymb}
\usepackage{amsmath}
\usepackage{float}
\usepackage{braket}
\usepackage{siunitx}
\usepackage{array}

\def\be{\begin{equation}}
\def\ee{\end{equation}}
\def \bea{\begin{eqnarray}}
\def \eea{\end{eqnarray}}
\def \nn{\nonumber}
\usepackage{amsmath}

\begin{document}

\title{Quantum Transport Spectroscopy of Pseudomagnetic Field in Graphene}

\author{Divya Sahani$^1$, Sunit Das$^2$, Kenji Watanabe$^3$, Takashi Taniguchi$^4$, Amit Agarwal$^2$}\author{Aveek Bid$^1$}
\email{aveek@iisc.ac.in, amitag@iitk.ac.in}
	\affiliation{$^1$Department of Physics, Indian Institute of Science, Bangalore 560012, India \\
		$^2$ Department of Physics, Indian Institute of Technology Kanpur, Kanpur 208016, India\\
        $^3$Research Center for Functional Materials, National Institute for Materials Science, 1-1 Namiki, Tsukuba 305-0044, Japan \\
		$^4$ International Center for Materials Nanoarchitectonics, National Institute for Materials Science, 1-1 Namiki, Tsukuba 305-0044, Japan\\}

\begin{abstract}

Nonuniform strain in graphene acts as a valley-dependent gauge field, generating pseudomagnetic fields (PMFs) that mimic real magnetic fields but preserve global time-reversal symmetry. While local probes have visualized such fields, their quantitative detection via macroscopic transport has remained elusive. Here, we demonstrate that high-mobility graphene exhibits distinct beating patterns in Shubnikov-de Haas oscillations, arising from valley-resolved Landau quantization under different effective magnetic fields. Systematic analysis of these beats reveals universal quadratic and linear scaling of the node carrier density and Landau level filling factor with the applied magnetic field, enabling the extraction of PMFs as small as a few millitesla. Our results establish quantum oscillation spectroscopy as a robust and broadly applicable probe of strain-induced gauge fields in Dirac materials, opening avenues for mechanically tunable valleytronic and straintronic devices.
\end{abstract}

\maketitle

{\it Introduction--} Mechanical deformation in two-dimensional materials provides a unique route to engineer emergent quantum phenomena~\cite{Miao_npj21, Pandey_23, Boland_acs24}. In graphene, nonuniform lattice strain couples to the Dirac Hamiltonian as a valley-dependent gauge potential ${\bm A}({\bm r})$, generating a pseudomagnetic field (PMF) ${\bm B}_{\rm pm} = {\bm \nabla} \times {\bm A} ({\bm r})$. These fields act with opposite polarity in the two inequivalent valleys, locally breaking time-reversal symmetry while preserving it globally. Theory predicts that such pseudomagnetic fields can reach hundreds of tesla in nanobubbles or engineered strain geometries~\cite{guinea2010energy, levy2010strain}. As a result, the Dirac spectrum acquires features analogous to those under real magnetic fields, including pseudo-Landau levels, valley-polarized edge modes, and valley-contrasting Berry phases~\cite{guinea2010energy, Guinea_nl10,vozmediano2010gauge,Mao2020,yoo2019, Kazmierczak2021, Halbertal2022, Zhu2015, Nigge2019, Guinea2010Ribbon,jia2019programmable, Liu_scd16, Reis_scd19,Kang_NC21, Bitan_prb23, Adak2024}. However, such extreme effects have so far been observed only in local probes such as scanning tunneling microscopy, or nano-optical imaging, which detect pseudo-Landau levels but cannot capture their consequences for bulk transport~\cite{Kang_NC21,levy2010strain, Lin_prb15, Lin_prl20, Shi_NC20, zhou2023imaging, Taniguchi_acs25, Mochales_NL25}. Macroscopic transport signatures remain elusive because strain-induced pseudomagnetic fields act with opposite polarity in the two valleys. Their contributions cancel, eliminating any net Hall response and rendering valley-antisymmetric gauge fields invisible in conventional transport.

Recent studies have associated thermal or geometry-induced deformations with signatures of pseudomagnetic field effects~\cite{Taniguchi_acs25,zhou2023imaging}. Still, a direct and quantitative detection of intrinsic strain-induced PMF through quantum transport in clean, extended graphene remains elusive. The mechanism by which such strain modifies global Landau quantization, quantum interference, and valley coupling in high-mobility graphene remains unclear. Addressing this question is crucial for realizing the mechanically reconfigurable functionalities of straintronics and valleytronics.

In this Letter, we report the quantitative detection of strain-induced PMF through quantum transport in high-mobility graphene. In the presence of an externally applied magnetic field $B$, the longitudinal resistance exhibits distinct beating patterns superimposed on Shubnikov-de Haas (SdH) oscillations. We show that these beats originate from the interference between valley-resolved Landau quantizations under the combined action of $B$ and the strain-induced pseudomagnetic field $B_{\rm pm}$. By mapping the scaling of beating nodes with magnetic field and carrier density, we extract pseudomagnetic fields as small as a few milliteslas, establishing SdH spectroscopy as a quantitative probe of emergent gauge fields in Dirac materials.

\begin{figure*}[t]
	\begin{center}
			\includegraphics[width=\linewidth]{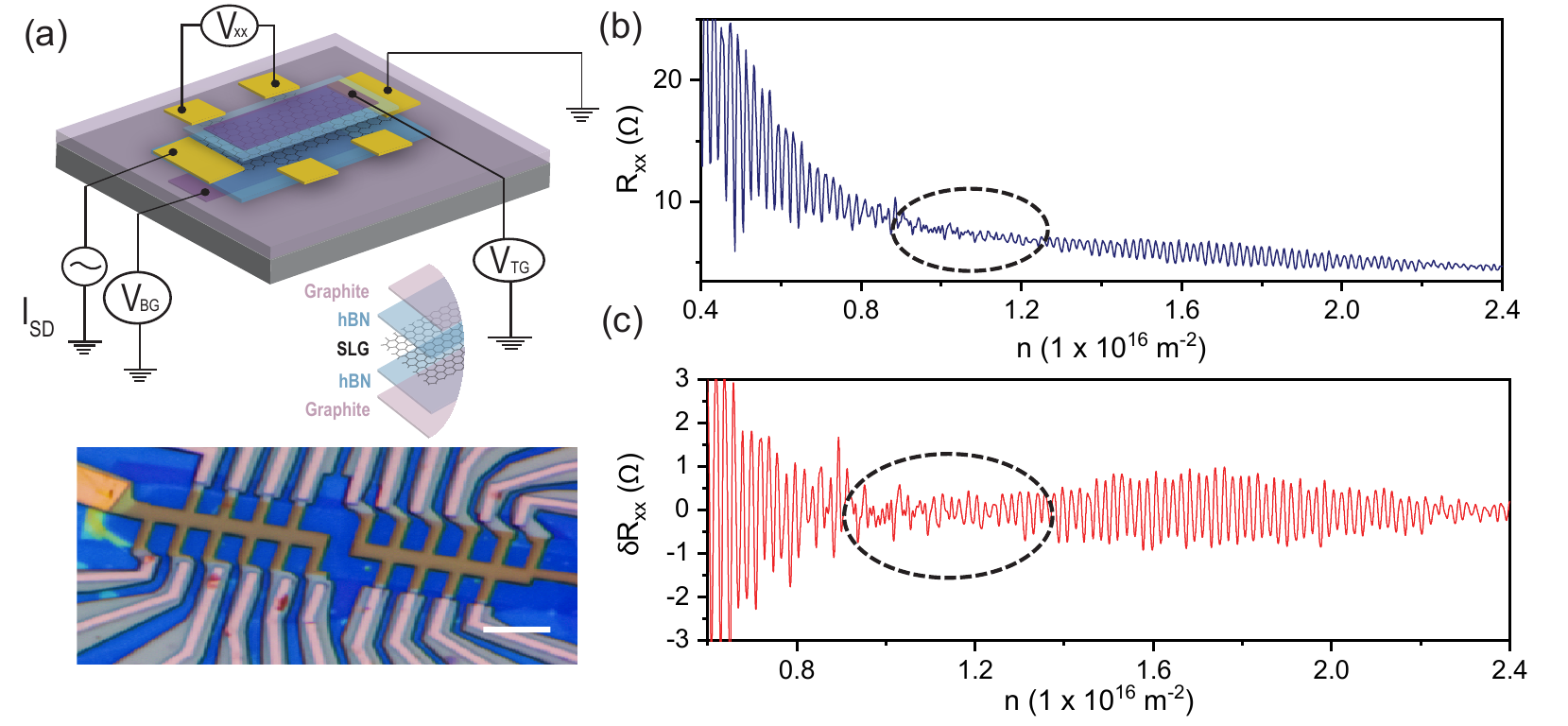}
			{\caption{\textbf{Device structure and amplitude modulation of quantum oscillation.}  {(a) Top panel: Schematic of the graphite/hBN/graphene/hBN/graphite dual-gated Hall bar device. Bottom panel: Optical image of the device. The scale bar is $3~\mathrm{\mu m}$. (b) Plot of longitudinal magnetoresistance $R_{xx}(B)$ as a function of carrier density $n$ at finite magnetic field  $B = 0.2$~T. (d) Plot of $\delta R_{xx}(B)$ versus $n$ in the presence of $B = 0.2$~T after removal of a smooth background. The regions marked by dashed ellipses in (b) and (c) mark nodes in the quantum oscillations.}
            \label{Fig:1}}}
        \end{center}
\end{figure*}
{\it Device and observation of SdH beating--} Dual graphite-gated graphene devices were fabricated using standard dry transfer techniques.  See Fig.~\ref{Fig:1}(a), and  Refs.~\cite{sahani2025giant,tiwari2023observation} for more details. In this Letter, we present the data from device D1, while data from other devices exhibiting similar characteristics are presented in the Supplemental Material (SM)~
\footnote{The Supplemental Material contains the details on (S1) Device characterization, (S2) oscillation data from device D2, (S3) the origin of in-built strain, (S4) calculation of PMF induced beating in magnetoconductivity, (S5) critical filling factor vs. magnetic field plot, (S6) the effect of parallel magnetic field in beating, (S7) estimation strain and strain gradient from PMF, (S8) extraction of PMF from Fast Fourier transform, and (S9) intrinsic valley polarization as the origin of beating.}. The device quality is exceptional, with charge-carrier mobility exceeding $\sim 3.5\times 10^5 \, \rm cm^2 V^{-1} s^{-1}$, as determined from zero-field transport (Section S1 of SM~\cite{Note1}). At a temperature of $20$~mK and a perpendicular magnetic field of $B=0.2$~T, well-resolved SdH oscillations appear in the longitudinal resistance $R_{xx}$ as a function of carrier density $n$ [Fig.~\ref{Fig:1}(b)], confirming extremely low disorder and long quantum lifetimes. Notably, the SdH oscillations exhibit a periodic suppression of amplitude, resulting in a well-defined beating envelope, as highlighted by the dashed ellipse in the plot. This feature, clearly visible after subtracting a smooth background [Fig.~\ref{Fig:1}(c)], indicates interference between two closely spaced SdH oscillation frequencies.

\begin{figure}[t!]
	\begin{center}
			\includegraphics[width=0.5\linewidth]{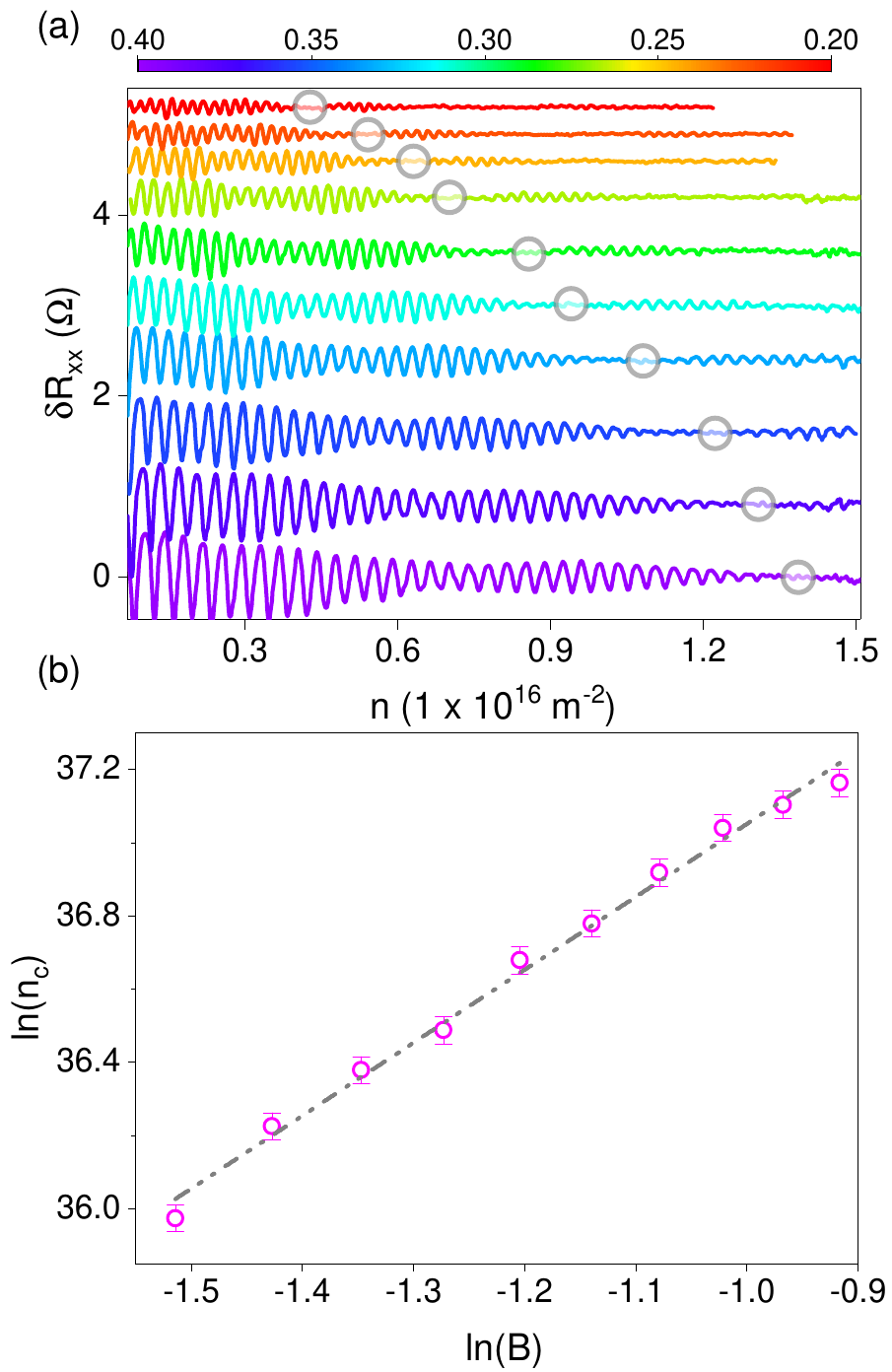}
			\caption{\textbf{Magnetic field dependent modulation of the beating nodes.} {(a) Plot of residual ${\delta R_{xx}}$ as a function of $n$ at different values of $B$ ranging from $\mathrm{0.22~T ~to~ 0.4~T}$  (indicated by the colorbar on the top). Each plot has been vertically offset by \SI{0.8}{\ohm} for clarity. (b) Plot of the node positions ln(${n_c}$) as a function of magnetic field ln($B$). The black dashed line shows the linear fit to the data with a slope of $\sim 1.99 \pm 0.06$, implying that $n_c \propto B^2$. }
            \label{Fig:2}}
        \end{center}
\end{figure}

{\it Magnetic field dependent modulation of beating--} To probe the mechanism behind the beating, we tracked the position of amplitude minima (`beating node') as a function of magnetic field. A plot of $\delta R_{xx}$ in Fig.~\ref{Fig:2}(a) as a function of carrier density $n$ over the range of $0.2~{\rm T} < B < 0.4~{\rm T}$ shows that the node positions shift systematically toward higher carrier density with increasing $B$: The gray circles in Fig.~\ref{Fig:2}(a) mark the the locus of this shift. Figure~\ref{Fig:2}(b) shows a plot of the carrier density corresponding to each node, ${\rm ln}(n_c)$ versus ${\rm ln}(B)$, yielding a slope of $1.99 \pm 0.06$. This scaling unambiguously shows that $n_c \propto B^2$. A similar analysis reveals that the associated Landau level filling factor ($\nu_c$) varies linearly with $B$ (Fig.~S3 of the SM). This trend is consistent with the observed increase in the critical carrier density $n_c$ with $B$, as the carrier density and filling factor are monotonically related in graphene due to its linear band dispersion. These two scaling relations, quadratic in $n_c$ and linear in $\nu_c$, hint at a fundamental connection between the underlying quantum interference mechanism and the Landau quantization in graphene.
These scaling relations are also satisfied by the data obtained from a second device (D2) (Section~S2 of the SM~\cite{Note1}). The high signal-to-noise ratio and reproducibility across multiple cooldowns and devices confirm that the observed beating is an intrinsic property of the graphene device, rather than due to device inhomogeneity or extrinsic artifacts.

\begin{figure*}[t]
    \centering
    \includegraphics[width=\columnwidth]{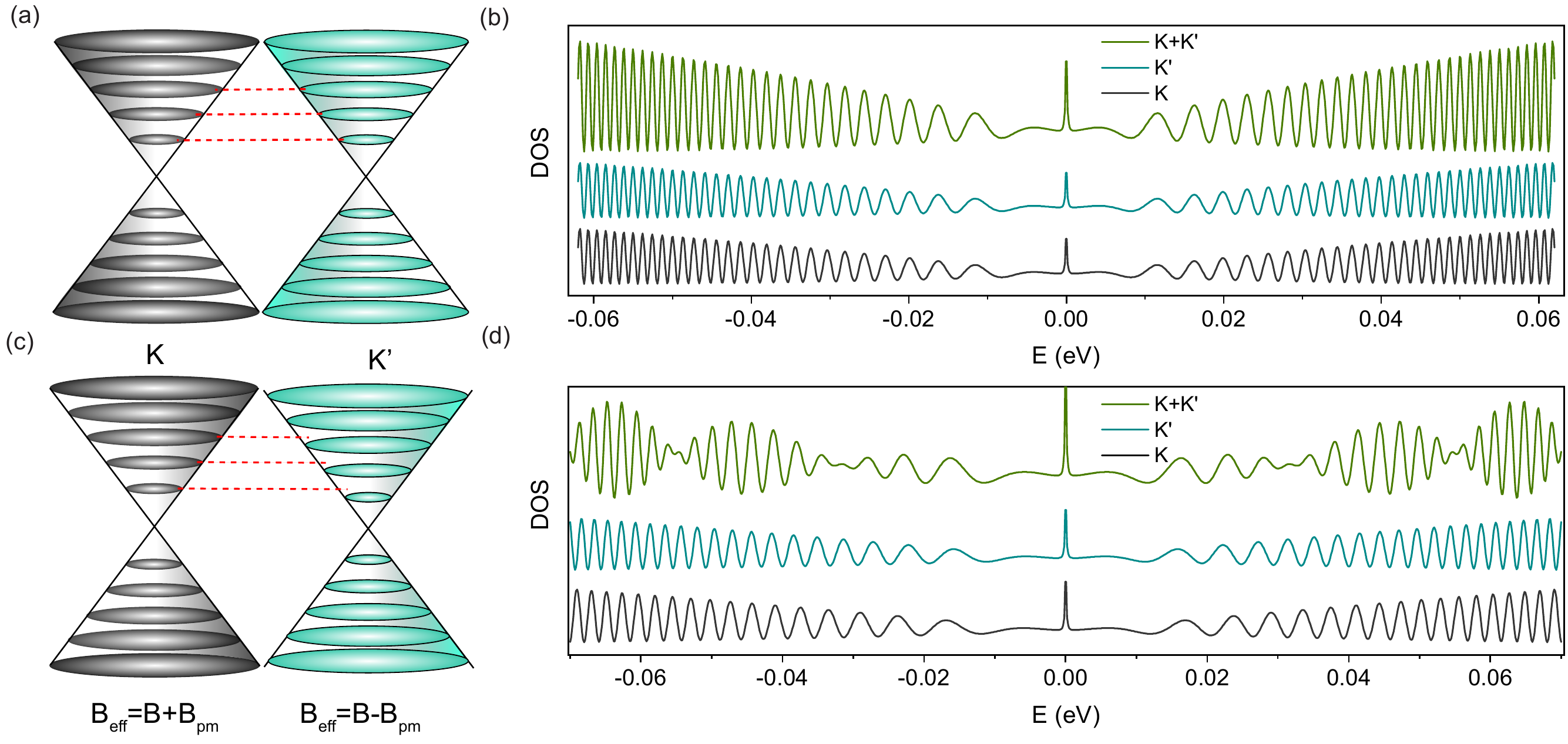}
    \caption{
    \textbf{Strain-induced pseudomagnetic field and the origin of beating in Shubnikov--de Haas oscillations.} (a) Schematic Landau level (LL) structure in single-layer graphene for the two valleys, $K$ and $K'$, under an external magnetic field $B$ in the absence of pseudomagnetic field ($B_{\mathrm{pm}} = 0$); both valleys are degenerate.  (b) Calculated normalized DOS for $K$, $K'$, and their combined contribution for $B = 0.2~\mathrm{T}$ and $B_{\mathrm{pm}} = 0$, showing conventional periodic SdH oscillations with no amplitude modulation. (c) Under a finite pseudomagnetic field $B_{\mathrm{pm}}$, charge carriers in the two valleys experience opposite effective fields $B_{\mathrm{eff}}^{K,K'} = B \pm B_{\mathrm{pm}}$, breaking valley degeneracy and inducing a relative phase shift between their LL spectra. (d) Calculated normalized DOS for $B = 0.2~\mathrm{T}$ and $B_{\mathrm{pm}} = 13.4~\mathrm{mT}$ showing clear amplitude modulation (beating) arising from valley interference between the shifted quantization ladders. The DOS plots in panels (b) and (d) are vertically offset for clarity. A constant LL broadening of $\sim 1$ meV has been used in all calculations. \label{Fig:3} }
\end{figure*}

{\it Pseudomagnetic field as the origin of beating--} We now turn to the microscopic origin of the beating pattern in SdH oscillations and the scaling of $n_c$ and $\nu_c$ with $B$. Beating in quantum oscillations typically reflects the presence of Fermi surface orbits of nearly equal areas at similar energies. In pristine single-layer graphene, the low-energy band structures near the two inequivalent valleys, $K$ and $K'$, are related by time-reversal symmetry. Consequently, the Fermi surface orbit areas and corresponding Landau quantization under a perpendicular field $B$ are strictly identical [Fig.~\ref{Fig:3}(a)], leading to perfectly phase-aligned DOS (and consequently, SdH) oscillations from the two valleys [Fig.~\ref{Fig:3}(b)].

Long-wavelength nonuniform mechanical strain (see Section~S3 of SM~\cite{Note1} for possible origin of strain) couples to the Dirac Hamiltonian as a valley-dependent vector potential ${\bm A}({\bm r})$, generating a PMF, ${\bm B}_{\rm pm} = {\bm \nabla} \times {\bm A} ({\bm r})$, acting in different direction in two valleys~\cite{vozmediano2010gauge, Nadya_apl19,hsu2020nanoscale}. Carriers in the $K$ and $K'$ valleys then experience different effective fields $B_{\mathrm{eff}}^{K, K'} = B \pm B_{\mathrm{pm}}$, which lift the valley degeneracy introducing a slight offset between their Landau level spectra [Fig.~\ref{Fig:3}(c)] and a relative phase shift between the valley-resolved oscillations of the DOS [Fig.~\ref{Fig:3}(d)]. The superposition of these slightly out-of-phase oscillations produces the periodic suppression of oscillation amplitude, i.e., beating pattern in total DOS, as shown in Fig.~\ref{Fig:3}(d) (analytical form in Section~S4 of the SM). Consequently, in a uniform external magnetic field, even a tiny PMF induces a valley-dependent cyclotron quantization, producing a detectable phase offset between the SdH oscillations of the two valleys. This phase offset manifests as the experimentally observed amplitude modulation, referred to as beating.

{\it Extraction of pseudomagnetic field--} As discussed above, the SdH oscillations in the two valleys are slightly detuned, and their interference modulates the observed quantum oscillations. To quantitatively analyze this behavior, we analytically calculate the valley-resolved oscillating magnetoconductivity in the presence of the effective magnetic fields $B \pm B_{\rm pm}$. The oscillating part of the total conductivity, obtained by summing the valley-resolved contributions, is given by
\be \label{sigma_osc}
\delta \sigma_{xx} \approx \sigma_0  \, \Omega_D \, \Omega_T \,   \cos\left(2\pi \frac{f_0(E)}{B}\right)  \cos\left(2\pi \frac{f_0(E) B_{\rm pm}}{B^2}  \right)~.
\ee
Here, $\sigma_0$ is the non-oscillatory background conductivity, $\Omega_D$ is the Dingle damping factor capturing disorder broadening, and $\Omega_T$ accounts for thermal smearing of the Landau levels. See Section S4 of SM~\cite{Note1} for explicit expressions of these quantities and the detailed derivation of $\delta \sigma_{xx}$. The first cosine term in Eq.~\eqref{sigma_osc}, periodic in $1/B$, represents the conventional SdH oscillation, with the frequency $f_0(E) =  E^2 / (2 e \hbar v_F^2)$.

\begin{figure*}
    \begin{center}
		\includegraphics[width=\columnwidth]{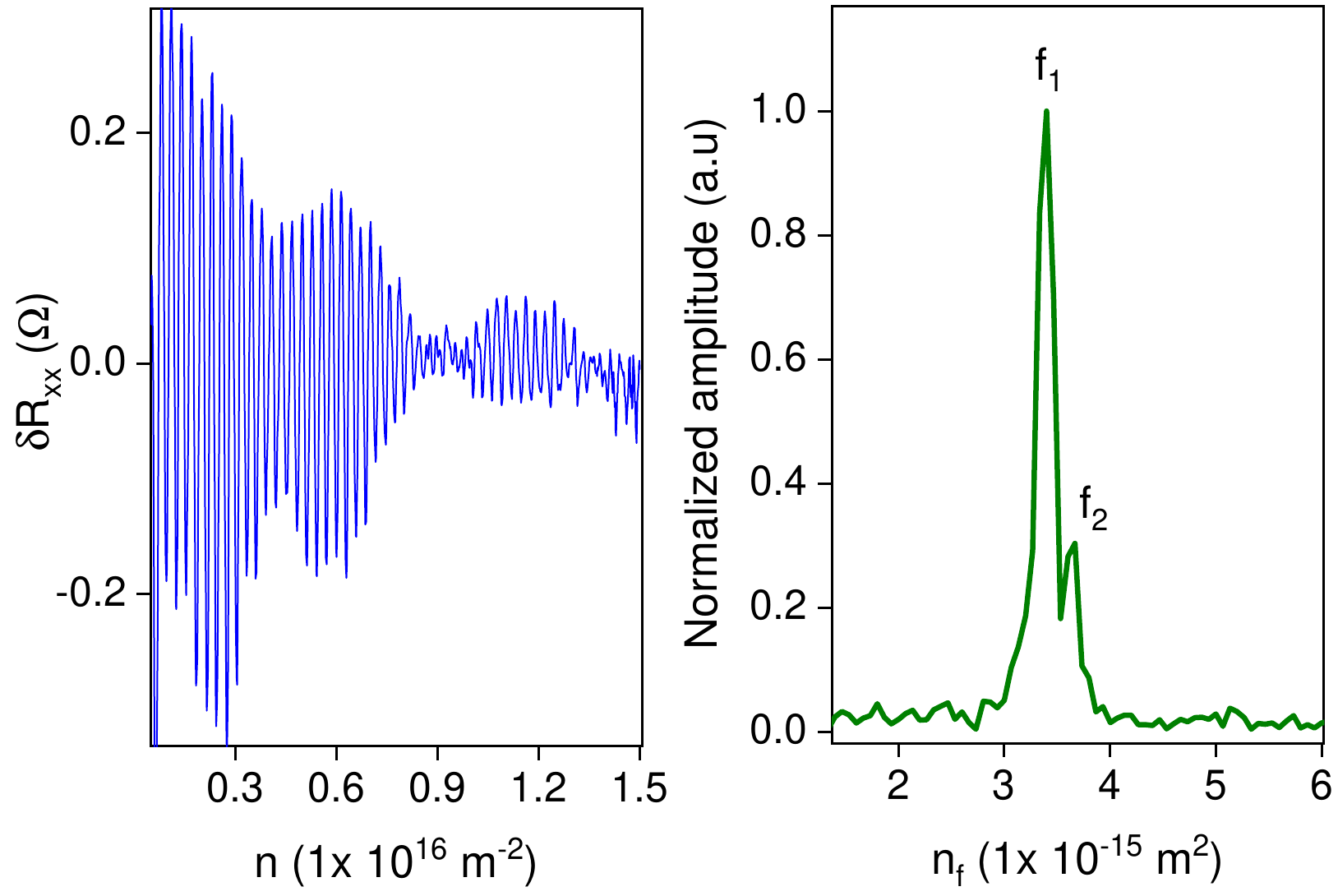}
		\caption{
    \textbf{Extraction of $B_{pm}$ from Fourier spectrum.} (a) Plot of $R_{xx}$ versus $n$ measured at $B=0.32$~T. (b) Fourier spectrum of the data in panel (a) showing the two Fourier peaks $f_1$ and $f_2$.
    \label{Fig:4}}
        \end{center}
\end{figure*}
In contrast, the second cosine term describes the valley interference envelope arising due to the PMF ($B_{\rm pm} \ll B$).  Beating nodes occur whenever this envelope term vanishes, i.e., $\cos \left( 2\pi \frac{f_0(E) B_{\mathrm{pm}}}{B^2} \right) = 0$. This yields the condition $E^2 B_{\rm pm}/( e \hbar v_F^2 B^2) = (2j + 1)/2$, with $j$ is an integer denoting the beating node index. For a linear Dirac band, $n = E^2 / (\pi \hbar^2 v_F^2)$. Consequently, we obtain the critical carrier density and the corresponding critical Landau level filling factor of the $j^{\rm th}$ beating node:
\begin{equation} \label{nc_cond}
n_c = \frac{eB^2}{h B_{\rm pm}}(2j + 1), \qquad
\nu_c = \frac{B}{B_{\rm pm}} {(2j + 1)}.
\end{equation}
Equation~\eqref{nc_cond} predicts two key experimental trends: $n_c \propto B^2$, and $\nu_c \propto B$. These can be understood as follows. With increasing $B$, the relative phase difference between the valley-resolved oscillations, which scales as $ E^2 B_{\mathrm{pm}}/B^2$, diminishes. Consequently, a higher Fermi energy (or carrier density) is required to fulfill the destructive-interference condition, naturally leading to the observed quadratic scaling $n_c \propto B^2$. The predicted dependencies are in quantitative agreement with our measurements (Fig.~\ref{Fig:2}(b) and Fig.~S3 of SM), establishing strain-induced PMFs as the microscopic origin of the observed beating in the magnetoresistance oscillations.

Fits to the $n_c$ versus $B$ plot in Fig.~\ref{Fig:2}(b) and $\nu_c$ versus $B$ plot in Fig.~S3 of SM~\cite{Note1} yield $B_{\mathrm{pm}} =13.4\pm 0.7$~mT. This best-fit value of $B_{\mathrm{pm}}$ corresponds to a characteristic strain gradient
$|\nabla \varepsilon| \approx 1.7 \times 10^{3}~\mathrm{m^{-1}}$; for micron-scale strain variations, this implies a local strain amplitude of order $10^{-3}$. This value is consistent with prior estimates for encapsulated graphene~\cite{guinea2010energy, vozmediano2010gauge} (see Section~S2 of SM~\cite{Note1} for details). The \textit{quantitative} detection of pseudomagnetic fields using bulk transport measurements is the central result of this Letter.

The beating pattern implies the presence of two nearby quantum oscillation frequencies. We obtain these by performing a Fourier transform of the beating pattern, which yields two well-resolved frequencies, $f_1$ and $f_2$ [Fig.~\ref{Fig:4}(a-b)]. Their separation directly encodes the pseudomagnetic field via (see Section~S8 of SM~\cite{Note1} for details)
\begin{equation}
B_{\rm pm} = B\frac{|f_1 - f_2|}{f_1 + f_2}.
\end{equation}
From the frequency splitting, we obtain $B_{\rm pm} = 12.0 \pm 0.6$~mT. This value is consistent, within experimental uncertainty, with that extracted from the node-scaling analysis, offering a complementary means to determine the pseudomagnetic field from a single magnetoresistance trace.

{\it Discussion and Conclusions--}  Amplitude modulation in the magnetoresistance could, in principle, arise from several mechanisms, including spin splitting~\cite{tiwari2023observation, Rao_nc23, hatke2012shubnikov}, multiple Fermi pockets, intrinsic valley polarization, sublattice asymmetry, Kekul\'e distortion, disorder, or magnetic breakdown. However, each of these predicts either field‐independent or non-quadratic node scaling and fails to reproduce the systematic carrier density and field dependence observed here (see End Matter for a detailed discussion). The invariance of the oscillation pattern under in-plane magnetic fields (Fig.~S4 of the SM~\cite{Note1}) further excludes spin–orbit and Zeeman contributions.  Thus, the only consistent interpretation is interference between valley-resolved Landau quantization under the effective fields $B \pm B_{\mathrm{pm}}$.

Finally, we comment on why the observed beating remains sharp despite the spatially varying strain present in encapsulated graphene. In hBN-encapsulated graphene, the pseudomagnetic field $B_{\mathrm{pm}}(\mathbf{r})$ is nonuniform but varies smoothly on a length scale much larger than the cyclotron radius in the magnetic-field range used in this work. Scanning-probe studies, including STM and nano-Raman measurements~\cite{levy2010strain, yoo2019, Kazmierczak2021, zeldov_nature23}, report characteristic strain correlation lengths of $200$--$800~\mathrm{nm}$. By contrast, at magnetic fields of $0.2$--$0.4$~T the cyclotron radius is only $r_c \approx 50$--$80~\mathrm{nm}$, so electrons sample an effectively uniform pseudomagnetic field along each cyclotron orbit. Consequently, the valley-dependent phase accumulated per orbit is well defined and survives spatial averaging. The beating originates from the \textit{relative} phase shift between the two valley-resolved Landau-level ladders, which experience opposite effective fields $B \pm B_{\mathrm{pm}}(\mathbf{r})$; this opposite polarity enforces a coherent valley-antisymmetric phase difference across the device. Spatial averaging therefore suppresses only the net valley-odd response (such as Hall voltages), while preserving the valley-phase interference responsible for the SdH beating. This hierarchy of length scales ($r_c \ll \xi_{\mathrm{strain}}$) explains the sharp, regularly spaced beating nodes and their robust quadratic--linear scaling, even in the presence of realistic strain inhomogeneity.

To conclude, our results demonstrate that quantum oscillations provide a quantitative probe for emergent gauge fields in two-dimensional Dirac materials. The beating in the Shubnikov-de Haas oscillations originates from valley-resolved Landau quantization under effective magnetic fields $B \pm B_{\mathrm{pm}}$, with the linear and quadratic scaling of $n_c$ and $\nu_c$ with $B$ serving as unambiguous transport fingerprints of valley-interference. The universal scaling enables millitesla-level sensitivity to pseudomagnetic fields, unifying local gauge-field phenomena with macroscopic transport. This framework offers a robust route toward strain-tunable valley polarization and reconfigurable quantum phases, opening pathways to strain-engineered valleytronic and quantum-sensing technologies.

\textbf{Acknowledgments:}
A.B. acknowledges funding from the U.S. Army DEVCOM Indo-Pacific (Project number: FA5209   22P0166) and ANRF, DST (Project number: SPR/2023/000185). K.W. and T.T. acknowledge support from the JSPS KAKENHI (Grant Numbers 21H05233 and 23H02052) and World Premier International Research Centre Initiative (WPI), MEXT, Japan.  S.D. acknowledges funding from the Prime Minister's Research Fellowship, Ministry of Education, Government of India.

{\it Note added--} During the final preparation of this manuscript, a related study reported a pseudomagnetic field in encapsulated graphene via quantum transport, attributed to the thermally generated strain gradient~\cite{Taniguchi_acs25}.

\clearpage

\begin{center}
{\textbf{\large Supplemental Materials: `Quantum Transport Spectroscopy of Pseudomagnetic Field in Graphene'}}
\end{center}

\newcommand{\cmark}{\ding{51}}
\newcommand{\xmark}{\ding{55}}
\renewcommand{\theequation}{S\arabic{equation}}
\renewcommand{\thesection}{S\arabic{section}}
\renewcommand{\thefigure}{S\arabic{figure}}
\renewcommand{\thetable}{S\arabic{table}}
\setcounter{table}{0}
\setcounter{figure}{0}
\setcounter{equation}{0}
\setcounter{section}{0}
 \tableofcontents
\section{Device characterization}
\begin{figure}[b]
	\begin{center}
\includegraphics[width=0.7\linewidth]{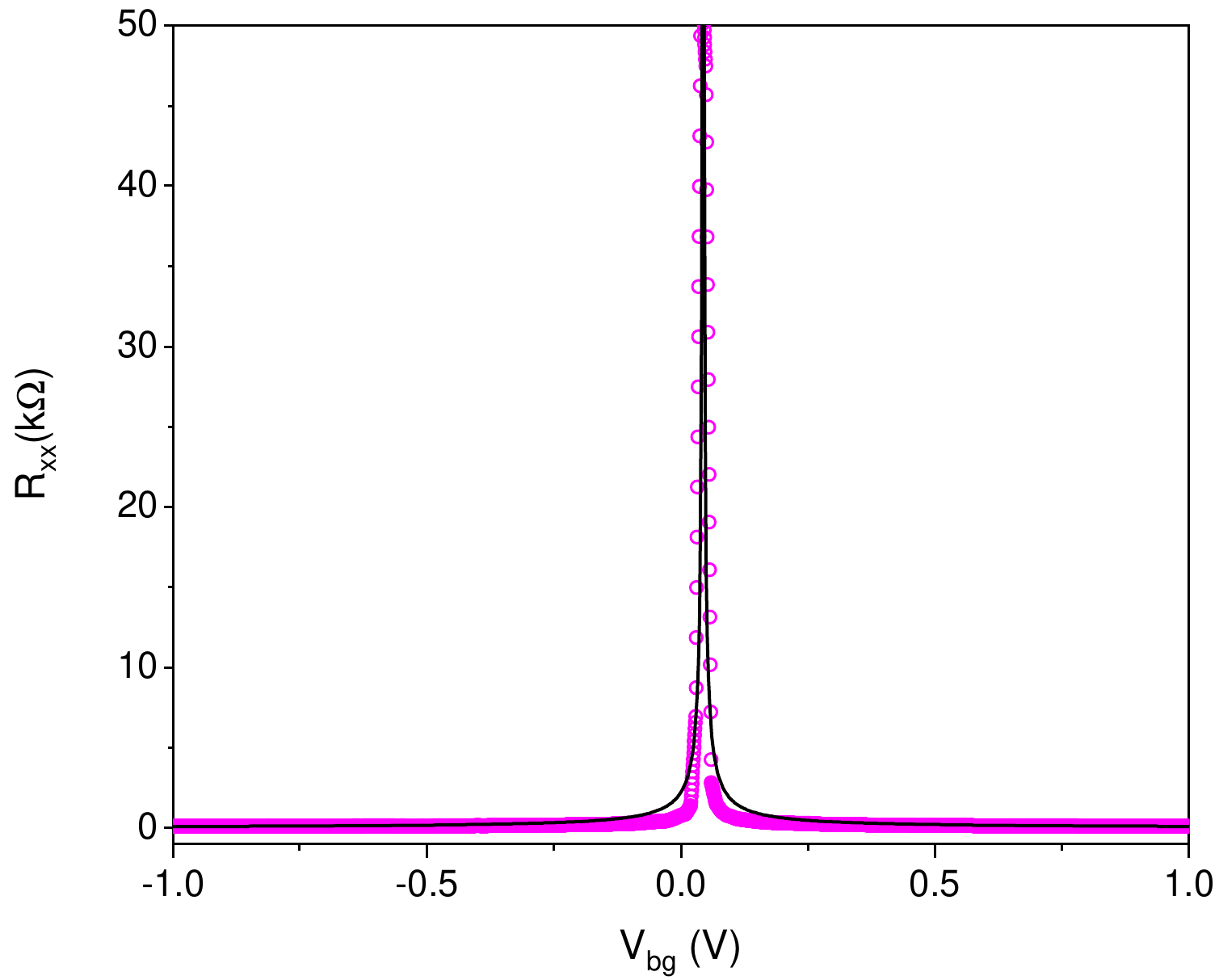}
			\small{\caption{\textbf{Device characterization.} {Plot of  $R_{xx}$ as a function of $V_{bg}$, measured at $T = 20~\mathrm{mK}$ and $B = 0~\mathrm{T}$ (pink open circles). The black line represents a fit to Eq.~\eqref{eg: S1}.}}\label{Fig:S1}}
    \end{center}
\end{figure}

To characterize the device, we measured the longitudinal resistance $R_{xx}$ as a function of back-gate voltage $V_{bg}$, at 20~$\mathrm{mK}$  as shown in Fig.~\ref{Fig:S1}. The black curve represents a fit to the data using the relation:
\begin{equation} \label{eg: S1}
R = R_c + \frac{L}{W e \mu \sqrt{n_0^2 + \left( \frac{C (V_{bg} - V_d)}{e} \right)^2 }};
\end{equation}
where $R_c$ is the contact resistance, $L$ and $W$ are the channel length and width, $\mu$ is the carrier mobility, $n_0$ is the residual carrier density, $C$ is the back-gate capacitance per unit area (extracted from Quantum Hall measurements), and $V_d$ is the Dirac point voltage. From the fit, we extract the mobility of the device to be $\sim 3.5\times 10^5 \, \rm cm^2 V^{-1} s^{-1}$.

\begin{figure}
	\begin{center}
			\includegraphics[width=0.9\linewidth]{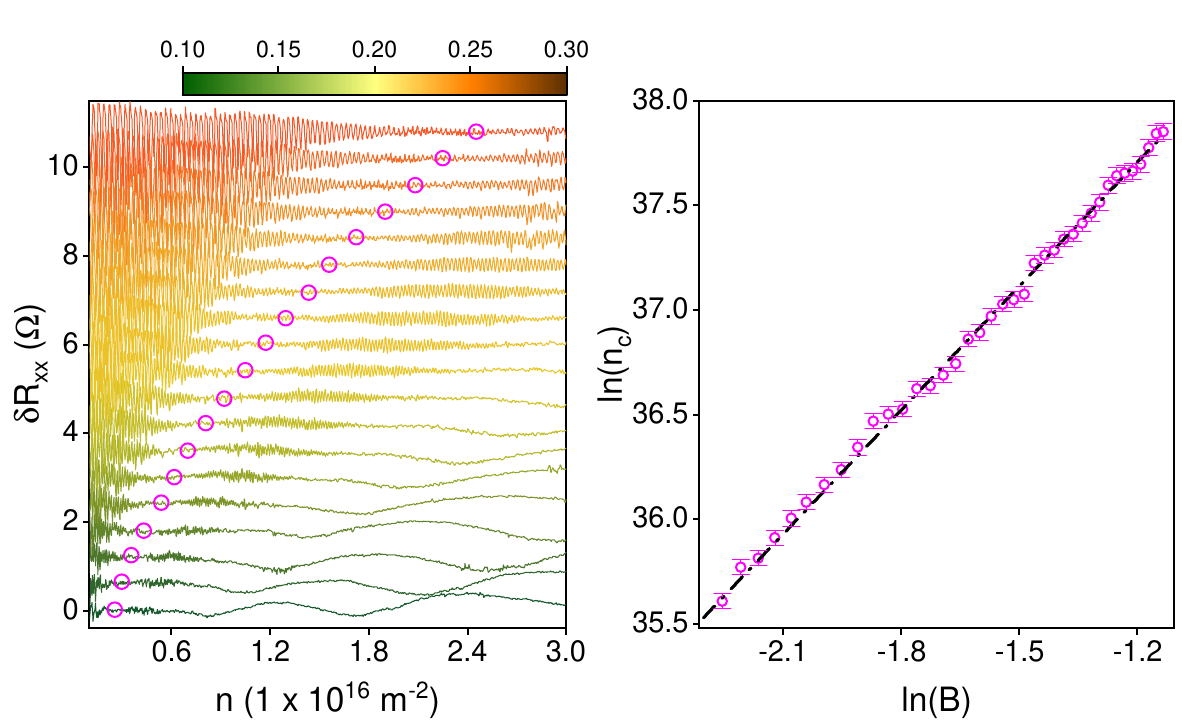}
			\small{\caption{\textbf{Data from device D2.} {(a) Plot of  ${\delta R_{xx}}$ as a function of $n$ at different values of $B$ ranging from $\mathrm{0.1~T ~to~ 0.3~T}$  (indicated by the colorbar on the top). (b) Plot of ${n_c}$ versus $B$ on a log-log scale. The black solid line is the linear fit with a slope $1.975 \pm 0.018$. } }\label{Fig:S2}}
        \end{center}
\end{figure}

 {\section{Data from device D2}}
Fig.~\ref{Fig:S2}(a) shows plots of $\delta R_{xx}$ versus $n$, of an additional device D2 measured over the magnetic field range $0.1~\mathrm{T} < B <0.3~\mathrm{T}$. We observe a similar behavior to that of device D1, where the nodes in $\delta R_{xx}$ shift toward higher $n$, a characteristic behavior of pseudomagnetic-field-induced beatings. To extract the magnitude of the pseudomagnetic field, we plot $n_c$ as a function of the applied magnetic field $B$ on a log-log scale in Fig.~\ref{Fig:S2}(b). From the linear fit, we obtain a slope of $1.975 \pm 0.018$, which confirms the expected quadratic relationship between $n_c$ and $B$. The intercept of the fit yields the amplitude of the pseudomagnetic field, which is $\approx 1.10~\mathrm{mT}$.

The two devices reported in this work exhibit different magnitudes of strain-induced pseudomagnetic field, with $B_{\mathrm{pm}} \approx 14~\mathrm{mT}$ for Device~D1 and $B_{\mathrm{pm}} \approx 1~\mathrm{mT}$ for Device~D2. Such variation is fully consistent with the known device-to-device variability of strain profiles in hBN-encapsulated graphene. During assembly, microscopic details such as polymer-assisted pickup, interfacial slippage, trapped nanobubbles, and differential thermal contraction during cooldown produce long-wavelength strain patterns whose gradients can differ by more than an order of magnitude between nominally similar devices. Scanning-tunneling and nano-Raman studies have reported pseudomagnetic fields ranging from sub-millitesla to several tens of millitesla depending on the local strain geometry and encapsulation conditions.

Despite the difference in $B_{\mathrm{pm}}$, both devices display the \emph{same} universal scaling relations predicted for valley-resolved Landau quantization under effective fields $B \pm B_{\mathrm{pm}}$: the critical node density follows $n_c \propto B^2$ and the corresponding critical filling factor satisfies $\nu_c \propto B$. This confirms that the underlying physical mechanism is identical in both devices, even though the absolute magnitude of the pseudomagnetic field depends sensitively on the specific strain landscape of each heterostructure. For completeness, strain-gradient estimates for both devices have been included in Section~S7.

 {\section{ Origin of inbuilt strain}}
Below, we list the possible sources of strain in the graphene channel. A summary of the various mechanisms is provided in Table~\ref{tab1}.

\begin{enumerate}

\item During the device fabrication using the dry transfer method, exfoliated graphene and hBN are sequentially picked up onto a polymer stamp, PC/PDMS, and then deposited onto a $\mathrm{SiO_2}$ substrate. During this process, non-uniform strain can develop due to deformation of the soft stamp, uneven mechanical forces during alignment, or temperature changes, resulting in a thermal mismatch between the polymer and the 2D layers \cite{ivanova2021mechanical,jain2018minimizing}.

\item Interfacial defects such as bubbles, wrinkles, or trapped residues can locally distort the graphene lattice, leading to strain within the graphene layer~\cite{levy2010strain,purdie2018cleaning,sanchez20212d}.

\item Slight bending or rippling of the graphene sheet during transfer or encapsulation can introduce out-of-plane deformation, further contributing to the overall strain field~\cite{pham2024transfer,zhou2023imaging,Guinea2010Ribbon}.

\item Another possible mechanism of strain in the graphene channel is a lattice mismatch of $\sim \mathrm{1.8~\%}$ between graphene and hBN. As graphene tries to conform to the underlying hBN lattice, its atomic structure experiences in-plane stretching or compression to accommodate the mismatch partially ~\cite{novoselov2014commensurate,tiwari2023observation}.

\item Thermal contraction mismatch between graphene, hBN, and gold contacts during cool down from high temperatures can generate a residual strain due to their differing thermal expansion coefficients. Overall, this results in lattice distortion, where atomic positions deviate from their equilibrium configuration, giving rise to strain within the graphene layer~\cite{Taniguchi_acs25,de2018strain}.

\end{enumerate}

\begin{table}[h]
	\centering
	\caption{Sources of strain in high-mobility hBN/graphene/hBN heterostructures.}
	\label{tab1}

	\begin{tabular}{|l| l| l|}
		\hline
		\textbf{Source of Strain} &
		\textbf{Physical Origin/Mechanism} &
		\textbf{Reference} \\
		\hline

		\parbox{4cm}{van der Waals adhesion \\ \& interfacial bubbles} &
		\parbox{7cm}{Local delamination or trapped hydrocarbons/water create dome-shaped regions under tension.} &
		\cite{Couto2014RandomStrain,Khestanova2016UniversalBubbles} \\
		\hline

		\parbox{4cm}{Wrinkles/folds} &
		\parbox{7cm}{Mechanical relaxation or buckling during transfer; shear or uniaxial strain along wrinkle lines.} &
		\cite{Deng2017WrinkleFree} \\
		\hline

		\parbox{4cm}{Thermal-expansion mismatch} &
		\parbox{7cm}{Different thermal expansion coefficients during annealing or cooldown.} &
		\cite{Yoon2011StrainDependentSplitting} \\
		\hline

		\parbox{4cm}{Polymer transfer \& peel-off forces} &
		\parbox{7cm}{Bending and adhesion gradients during PPC/PMMA release.} &
		\cite{Zomer2014FastPickUp} \\
		\hline

		\parbox{4cm}{Substrate roughness} &
		\parbox{7cm}{Conformation to hBN or SiO$_2$ topography.} &
		\cite{Ishigami2007AtomicStructure} \\
		\hline

		\parbox{4cm}{Moiré mismatch \& alignment strain} &
		\parbox{7cm}{Lattice mismatch → moir\'e-periodic strain near alignment.} &
		\cite{Yankowitz2012SuperlatticeDirac,woods2014commensurate} \\
		\hline

		\parbox{4cm}{Electrostatic gating / displacement field} &
		\parbox{7cm}{Gate-induced pressure imbalance or piezoelectric effects.} &
		\cite{Freitag2016ElectrostaticallyConfined} \\
		\hline

		\parbox{4cm}{Metal contact stress} &
		\parbox{7cm}{Thermal expansion and intrinsic film stress during evaporation.} &
		\cite{Taniguchi_acs25} \\
		\hline

		\parbox{4cm}{Etching and patterning} &
		\parbox{7cm}{Edge stress and strain relaxation.} &
		\cite{Chae2012Renormalization} \\
		\hline

		\parbox{4cm}{Cryogenic cooldown strain} &
		\parbox{7cm}{Differential contraction during cooldown.} &
		\cite{Dean2010BoronNitride} \\
		\hline

	\end{tabular}
\end{table}

{\section{Pseudomagnetic field induced beating in oscillating density of states and magnetoconductivity}} \label{Sec_S4}
In this Section, we calculate the Landau levels, the oscillating density of states, and the oscillating magnetoconductivity (Lifshitz-Kosevich formula) for graphene. The low-energy model Hamiltonian for graphene is given by
\be
{\cal H}_\xi = \hbar v_F ( \xi \sigma_x k_x + \sigma_y k_y) ,
\ee
where $\xi$ is the valley index. In the presence of a magnetic field along the $z$-direction, the Hamiltonian becomes
\be
{\cal H}_B =  v_F \big[ \xi (\hbar k_x +eA_x) \sigma_x + (\hbar k_y +eA_y) \sigma_y) \big] .
\ee
To evaluate the Landau levels in the presence of the out-of-plane magnetic field, we choose the Landau gauge, ${\bm A} =  (0,xB,0)$. The Landau levels are obtained to be~\cite{Sunit_25}
\begin{subequations}
\bea
E_N &=&  \lambda \sqrt{2N (\hbar \omega_c)^2}, ~~~~ \text{for $N \neq 0$}, \\
E_0 &=& 0 ~~~~~~~~~~~~~~~~~~~~~\text{for $N = 0$.}
\eea
\end{subequations}
Here, $\lambda = \pm 1$ represents the band index. $\omega_c = v_F/l_B$ is the cyclotron frequency, with the magnetic length $l_B=\sqrt{\hbar/eB}$. The corresponding eigenstates are,
\bea
   \Phi_{N,k_y}^{\xi =+1}(X) = \frac{e^{i k_y y}}{\sqrt{2 L_y}}
\begin{pmatrix}
     \varphi_{N-1}(X) \\
    i \varphi_N(X)
\end{pmatrix}, ~\text{and} ~\Phi_{N,k_y}^{\xi = -1}(X) = \frac{e^{i k_y y}}{\sqrt{2 L_y}}
\begin{pmatrix}
    \varphi_{N}(X) \\
    i \varphi_{N-1}(X)
\end{pmatrix},
\eea
where $\varphi_N(x)$ is the Harmonic oscillator wavefunction. $X = (x + x_0) / l_B$ is the dimensionless position operator with the center of cyclotron orbit at $x = -x_0 = -k_y l_B^2$.

{\subsection{Oscillating DOS calculation}}
In an ideal, non-interacting system, the density of states (DOS) consists of sharp delta-function peaks at the Landau level energies. However, in realistic materials, these levels are broadened due to scattering from impurities, phonons, and electron-electron interactions. Such broadening and the resulting damping of quantum oscillations cannot be captured in a purely non-interacting framework. Instead, they are incorporated through the self-energy in the Green's function formalism. Self-energy modifies the single-particle Green's function, introducing energy shifts and finite lifetimes for quasiparticles. These effects are encoded in the spectral function, which describes how the spectral weight is distributed and broadened across energies. Integrating this spectral function over momentum yields the DOS, now modified to reflect interaction-induced effects.
Consequently, we calculate the oscillating density of states for the Landau levels from the self-energy. The self-energy in the presence of impurities is given by
\be
\Sigma(E) = \Gamma_0^2 \sum_N \frac{1}{E - E_N - \Sigma(E)},
\ee
where $\Gamma_0^2 $ characterizes the impurity scattering strength, and the sum is over all energy levels. The density of states is related to the self-energy via the following relation (including spin degeneracy)~\cite{Zhang_prb90}
\bea \label{DOS_self-energy}
D(E) = \frac{2}{2\pi l_B^2}{\rm Im}\left[\frac{\Sigma (E)}{\pi \Gamma_0^2} \right].
\eea
Note that $\frac{2}{2\pi l_B^2} = 2\frac{eB}{h}$ accounts for the degeneracy of each Landau level.

The self-energy is given by
\bea
\Sigma(E)= \sum_{N=0}^\infty \frac{1}{E - \lambda \sqrt{2 N (\hbar \omega_c)^2 } - \Sigma(E)} .
\eea
This sum is self-consistent due to \( \Sigma^-(E) \) appearing on both sides. The contribution of the zeroth Landau level to the DOS is given by $D(E) = \frac{1}{(\pi l_B \Gamma_0)^2} {\rm Im} [\frac{1}{E -\Sigma}]$.

We'll evaluate it using the residue theorem. We define
\bea
f(N) = \frac{1}{E - \sqrt{2 N (\hbar \omega_c)^2} - \Sigma(E)}.
\eea
To evaluate the sum over $N$, we use the contour integral identity
\bea
\sum_{N=0}^\infty f(N) = -\frac{1}{2\pi i} \oint_C f(z) \pi \cot(\pi z) \, dz,
\eea
where \( C \) encloses the non-negative integers $N = 0, 1, 2, \cdots$. Replacing $N$ with a complex variable \( z \), we obtain
\bea
f(z) = \frac{1}{E - \sqrt{2 z (\hbar \omega_c)^2 } - \Sigma(E)},
\eea
whose poles are given by $z_0 = \frac{[E -\Sigma(E)]^2}{2 (\hbar \omega_c)^2}$. Consequently, we have
\bea
\sum_N \frac{1}{E - E_N - \Sigma(E)} = - {\rm Res}[f(z)\pi \cot(\pi z)] &=&- \left. \frac{\pi \cot(\pi z)}{\frac{d}{dz} (E -E_z - \Sigma(E) )} \right|_{z_0} = -\frac{\pi \cot(\pi z_0) E_{z_0}}{-(\hbar \omega_c)^2} \nn \\
&=& \frac{\pi [E -\Sigma(E)] }{(\hbar \omega_c)^2} \cot\left(\pi \frac{[E -\Sigma(E)]^2 }{2 (\hbar \omega_c)^2} \right).
\eea
The self-consistent expression of the self-energy is given by
\be
\Sigma(E) \approx \frac{ \pi \Gamma_0^2 E}{(\hbar \omega_c)^2}  \cot(\pi N_0) ,
\ee
with $N_0=u+iv$. The $u$ and $v$ are given by
\bea
u = \text{Re}(N_0) = \frac{(E - \Sigma_r)^2 - \Sigma_i^2 - \Delta^2}{2 (\hbar \omega_c)^2}, ~~v = \text{Im}(N_0) = \frac{ (E - \Sigma_r) \Sigma_i}{ (\hbar \omega_c)^2}.
\eea
Since $\Sigma(E)$ is a complex quantity, we can write it as $\Sigma(E) = \Sigma_r + i \Sigma_i$. Furthermore, we have $\cot(\pi(u+i v)) = \dfrac{\sin(2\pi u) + i \sinh(2\pi v)}{\cos(2\pi u) + \cosh(2\pi v)}$. Using these, we obtain the imaginary part of the self-energy as follows
\be
{\rm Im}[\Sigma(E)]  \approx   \frac{ \pi \Gamma_0^2 E}{(\hbar \omega_c)^2}  \frac{\sinh(2\pi v)}{\cos(2\pi u) + \cosh(2\pi v)} = \frac{ \pi \Gamma_0^2 E}{(\hbar \omega_c)^2} \left[1+ 2 \sum_{p=1}^{\infty} e^{-p 2\pi v} \cos(2\pi p u) \right].
\ee
Here, we have used the relation $\dfrac{\sinh 2\pi v}{\cos 2\pi u + \cosh 2\pi v} = 1 + 2 \sum_{p=1}^\infty e^{-p 2\pi v} \cos (p 2\pi u)$, $p$ denoting the different harmonics. In the limit, $E \Sigma_i \ll (\hbar \omega_c)^2$, the $\Sigma_i$ can be obtained iteratively. The first iteration provides us $\Sigma_i = \pi \Gamma_0^2 E/(\hbar \omega_c)^2$. Consequently, using the above equation and Eq.~\eqref{DOS_self-energy}, we obtain the expression of the density of states as follows (neglecting energy renormalization terms due to $\Sigma$)
\bea \label{DOS_osc_final}
D(E) & \approx & \frac{E}{\pi(  \hbar v_F)^2} \left[1+ 2 \sum_{p=1}^{\infty} \exp\left(- 4 \pi^2 p \frac{ (\Gamma_0 E)^2}{(\hbar\omega_c)^4}  \right) \cos\left(2\pi p \frac{E^2}{2 (\hbar \omega_c)^2}\right) \right] . \label{DOS_osc}
\eea
The Eq.~\eqref{DOS_osc} presents the oscillating density of states due to the Landau levels, with oscillation frequency $f_0(E) = E^2 /(2\hbar^2 \omega_c^2 )= E^2/(2e\hbar v_F^2)$. It also incorporates the effect of impurity broadening via the damping term $\Omega_D = \exp\left(- 4 \pi^2 p \frac{ (\Gamma_0 E)^2}{(\hbar\omega_c)^4}  \right)$. We have derived the DOS for the conduction band, $\lambda=+1$. Nonetheless, a straightforward, similar calculation for $\lambda=-1$ can be performed. The DOS expression can be obtained by replacing $E$ with $|E|$, implying that the DOS is symmetric on both the conduction and valence band sides. This is physically consistent since pristine graphene does not break the particle-hole symmetry; the DOS should be similar on both the electron and hole sides.

{\subsection{Oscillating magnetoconductivity calculation}}
Having derived an oscillating DOS for graphene, we now derive the magnetoconductivity expressions. In a two-dimensional system under a strong perpendicular magnetic field, energy levels become quantized into perfectly flat Landau bands, leading to zero group velocity and, ideally, no longitudinal conductivity. However, in real materials, impurities and disorder broaden these levels, creating extended states where electrons can scatter between Landau orbits. This scattering, driven by charge impurities, gives rise to collisional conductivity, which governs longitudinal magneto-conductivity at low temperatures. The collisional conductivity in this regime can be evaluated using the following formula~\cite{Vasil_jmp82, Vasil_prb92}
\bea \label{coll_conduct}
    \sigma^{\text{col}}_{xx} = \frac{\beta e^2}{2 S_0} \sum_{\zeta,\zeta'} f_\zeta \left(1 - f_{\zeta'}\right) W_{\zeta \zeta'} \left(x_\zeta - x_{\zeta'}\right)^2.
\eea
Here, the $\zeta$ represents a quantum state, including all the quantum indices $\{\xi,\lambda,N,k_y\}$, and $S_0$ is the area of the sample. The average value of the $x$-component of the position operator of an electron in a particular quantum state is given by $x_\zeta = \bra{\zeta} x \ket{\zeta} = -k_y l_B^2$.

The scattering rate between states $\zeta$ and $\zeta'$ is given by,
\begin{equation}\label{scat rate}
    W_{\zeta \zeta'} = \frac{2\pi n_{\rm im}}{S_0 \hbar} \sum_q |U_q|^2 |F_{\zeta \zeta'}(\eta)|^2 \delta(E_\zeta - E_{\zeta'}) \delta_{k_y, k'_y + q_y},
\end{equation}
where $n_{\rm im}$ is the impurity density. The Fourier transformation of the screened charged impurity potential $U(r) = \left[\frac{e^2}{4\pi \epsilon_0 \epsilon_r r}\right] e^{-k_s r}$ is given by $U_q = U_0 [q^2 + k_s^2]^{-1/2} \approx \frac{U_0}{k_s}$ under the limit of small $ |q| \ll k_s $ for short-range Delta function-like potential. Here, $r$ is the real space distance from the impurity, $\epsilon_0$ is free space permittivity, and $\epsilon_r$ denotes relative permittivity. Here, $U_0 = \frac{e^2}{2 \epsilon_0 \epsilon_r}$ and $k_s$ is the screening vector. The function $F_{\zeta \zeta'}(\eta)=\bra{\zeta}  e^{i{\bm q}\cdot{\bm r}}\ket{\zeta'}$ denotes the form factor with its argument being $\eta = q^2 l_B^2 / 2$. The form factor is defined as,
\bea
    F_{\zeta \zeta'} = \langle \zeta | e^{i {\bm q} \cdot {\bm r}} | \zeta'\rangle = \langle\Phi_{N, k_y} | e^{i {\bm q} \cdot {\bm r}} | \Phi_{N', k_y'} \rangle.
\eea
Using these, the magnetoconductivity expression for a gapped Dirac model was recently derived in Ref.~\cite{Sunit_25}. The oscillating magnetoconductivity for valley $\xi$ is found to be (see Ref.~\cite{Sunit_25})
\bea \label{sigma_n}
\sigma_{xx}^{\xi} = \frac{e^2}{h}\frac{n_{\rm im} U_0^2}{ \pi k_s^2 l_B^2 \Gamma_0} \sum_{N} N \delta(E_N - E_F).
\eea
The above represents the oscillating magnetoconductivity contributed by the $N\neq 0$ Landau levels. The zeroth Landau level's contribution to the conductivity is given by
\be
\sigma_{xx}^{\xi} = \frac{e^2}{h}\frac{n_{\rm im} U_0^2}{2\pi k_s^2 l_B^2 \Gamma_0} \delta(E_F).
\ee
Note that the zeroth Landau level does not contribute to the oscillation.

To obtain the analytical form of the oscillating magnetoconductivity of Eq.~\eqref{sigma_n}, we can replace the summation over $N$ as follows~\cite{Islam_jpcm14_thermo}: $\sum_N \to 4\pi l_B^2 \int D(E) dE$, with $D(E)$ being the oscillating density of states. At finite temperature, the approximate expression of the conductivity for valley $\xi$ becomes
\be \label{sigma_osc_final}
\sigma_{xx}^\xi \approx \frac{e^2}{h}\frac{n_{\rm im} U_0^2}{4\pi^2 k_s^2 l_B^2 \Gamma_0} \frac{E_F^3}{ ( \hbar^2 v_F \omega_c)^2 } \left[1+ 2 \sum_{p=1}^{\infty} \Omega_T \exp\left(- 4 \pi^2 p \frac{E_F^2 \Gamma_0^2}{(\hbar\omega_c)^4}\right)  \cos\left(2\pi p \frac{E_F^2}{2 (\hbar \omega_c)^2}\right) \right] .
\ee
Here, $\Omega_T = \dfrac{T/T_D}{\sinh(T/T_D)}$ is the temperature-dependent broadening factor, with the Dingle temperature $T_D = e\hbar v_F^2 B/ (2\pi E_F k_B)$. The above equation is the Lifshitz-Kosevich formula for oscillating magnetoconductivity in graphene.

{\subsection{Beating in quantum oscillation}}

The Eq.~\eqref{DOS_osc_final} and Eq.~\eqref{sigma_osc_final} represent the oscillation of the density of states and the longitudinal magnetoconductivity for each valley experiencing the same externally applied magnetic field. We incorporate the effect of strain-induced opposite pseudomagnetic field in the valleys, by replacing $B$ with $B+\xi B_{\rm pm}$. Then the total DOS becomes (considering only the first harmonic, $p=1$)
\be \label{DOS_add}
D(E) = \frac{2E}{\pi( \hbar v_F)^2} \left[ 1 + \exp\left(- 4\pi^2 \frac{E_F^2 \Gamma_0^2}{(\hbar\omega_c)^4}\right) \left\{ \cos\left(2\pi \frac{f_0(E)}{B + B_{pm}} \right) + \cos\left(2\pi \frac{f_0(E)}{B - B_{pm}} \right) \right\} \right],
\ee
where $f_0 = E^2/(2\hbar e v_F^2)$. We have used the fact that the pseudo-magnetic field modifies the effective magnetic field in each valley as $B_\xi = B + \xi B_{\rm pm}$ with $\xi = \pm$. For $B_{\rm pm}/B \ll 1$, we can neglect the effect of $B_{\rm pm}$ on the damping factors.

The above DOS can be rewritten as
\bea \label{DOS_beating}
D(E, B) &\approx& D_0(E) \left[ 1 + 2 \Omega_D \cos\left(2\pi \frac{f_0(E)}{B} \right) \cos\left(2\pi \frac{f_0(E) B_{pm}}{B^2} \right) \right],
\eea
where $D_0(E) = 2E/(\pi \hbar^2 v_F^2)$ is the smooth monotonic part of the total DOS, and we denote $\Omega_D= \exp\left(- 4\pi^2 \frac{E_F^2 \Gamma_0^2}{(\hbar\omega_c)^4}\right)$. The above expression has been used to plot the density of states in Fig.~3 in the main text, where we choose $\Gamma_0\approx 1$ meV, and $B=0.2$ T.

Similarly, the total magnetoconductivity is obtained to be
\be \label{sigma_beating}
\sigma_{xx}^\xi \approx \frac{e^2}{h}\frac{n_{\rm im} U_0^2}{2\pi^2 k_s^2 l_B^2 \Gamma_0} \frac{E_F^3}{ ( \hbar^2 v_F \omega_c)^2 } \left[1+ 2 \Omega_T \Omega_D \cos\left(2\pi \frac{f_0(E)}{B} \right) \cos\left(2\pi \frac{f_0(E) B_{pm}}{B^2} \right) \right].
\ee
Both the DOS [Eq.~\eqref{DOS_beating}] and conductivity [Eq.~\eqref{sigma_beating}] contain a smoothly varying non-oscillating term and an oscillating term that consists of two cosine terms. We write the oscillating part of the conductivity as
\be \label{delta_sigma}
\delta \sigma_{xx} \approx \sigma_0  \, \Omega_D \, \Omega_T \,   \cos\left(2\pi \frac{f_0(E)}{B}\right)  \cos\left(2\pi \frac{f_0(E) B_{pm}}{B^2}  \right)~,
\ee
where $\sigma_0 = \frac{e^2}{h}\frac{2 n_{\rm im} U_0^2}{2\pi^2 k_s^2 l_B^2 \Gamma_0} \frac{E_F^3}{ ( \hbar^2 v_F \omega_c)^2 } $. The first cosine term in Eq.~\eqref{delta_sigma} is periodic in $1/B$, with conventional SdH oscillation with frequency $f_0(E)$. The second cosine term is aperiodic in $1/B$~\cite{Sunit_24_NL} and exhibits a slower oscillation frequency, and indicates the presence of systematic suppression of oscillation amplitude---the appearance of beating nodes. In Fig.~4(b) of the main text, we plotted the ${\rm ln} (\delta \sigma_{xx})$ in the $B^2-n$ space, where $n = E^2 /(\pi \hbar^2 v_F^2)$ is the carrier density of linear Dirac bands. We used the following parameters: $\Gamma_0 \approx 1$ meV, $v_F \approx 10^6$ m/s, $T=2$ K, $B_{\rm pm}=13.4$ mT, $k_s \sim 10^8$ m$^{-1}$.

The beating nodes occur when this envelope function vanishes, i.e.,
\be
\cos\left(2\pi \frac{f_0(E) B_{\rm pm}}{B^2} \right) = 0.
\ee
This condition yields
\be
2 \pi \frac{E^2 B_{\rm pm}}{2 \hbar e v_F^2 B^2} = \frac{2j + 1}{2} \pi.
\ee
Here, $j=0,1,2, \cdots$ denotes the beating node index. Now, using the expression $n = \dfrac{E_F^2}{\pi \hbar^2 v_F^2}$, the relation between the carrier density and the $j$-th beating node is obtained to be
\bea \label{nc_cond1}
n_c = \frac{e B^2}{ h B_{\rm pm}} (2j+1).
\eea
The above expression reveals a universal trend for monolayer graphene: the number density corresponding to the beating nodes varies quadratically with the applied magnetic field. More importantly, by finding the node position $j$, the $n$ vs. $B^2$ plot provides estimates of the strain-induced pseudomagnetic field present in the sample. Using the relation $\nu = n\dfrac{h}{e B}$, the condition in Eq.~\eqref{nc_cond1} can be rewritten as
\be
\nu_c = \frac{B}{B_{\rm pm}} (2j+1).
\ee
This provides the relation between the critical filling factor and the applied magnetic field.
Furthermore, using the Landau level expression $E^2 = 2N (\hbar \omega_c)^2$ in the above equation, we obtain the Landau level index corresponding to the $j^{\rm th}$ beating node:
\be
N_c =  \frac{B}{4 B_{\rm pm}} (2j+1).
\ee
This indicates that the critical Landau level index increases linearly with the applied magnetic field. The first beating node $j=0$ occurs for $N_c = B/4B_{\rm pm}$, consistent with Ref.~\cite{zeldov_nature23}.

\section{ The $\nu_c$ versus $B$ plot of device D1}
\begin{figure}
    \begin{center}
			\includegraphics[width=0.5\linewidth]{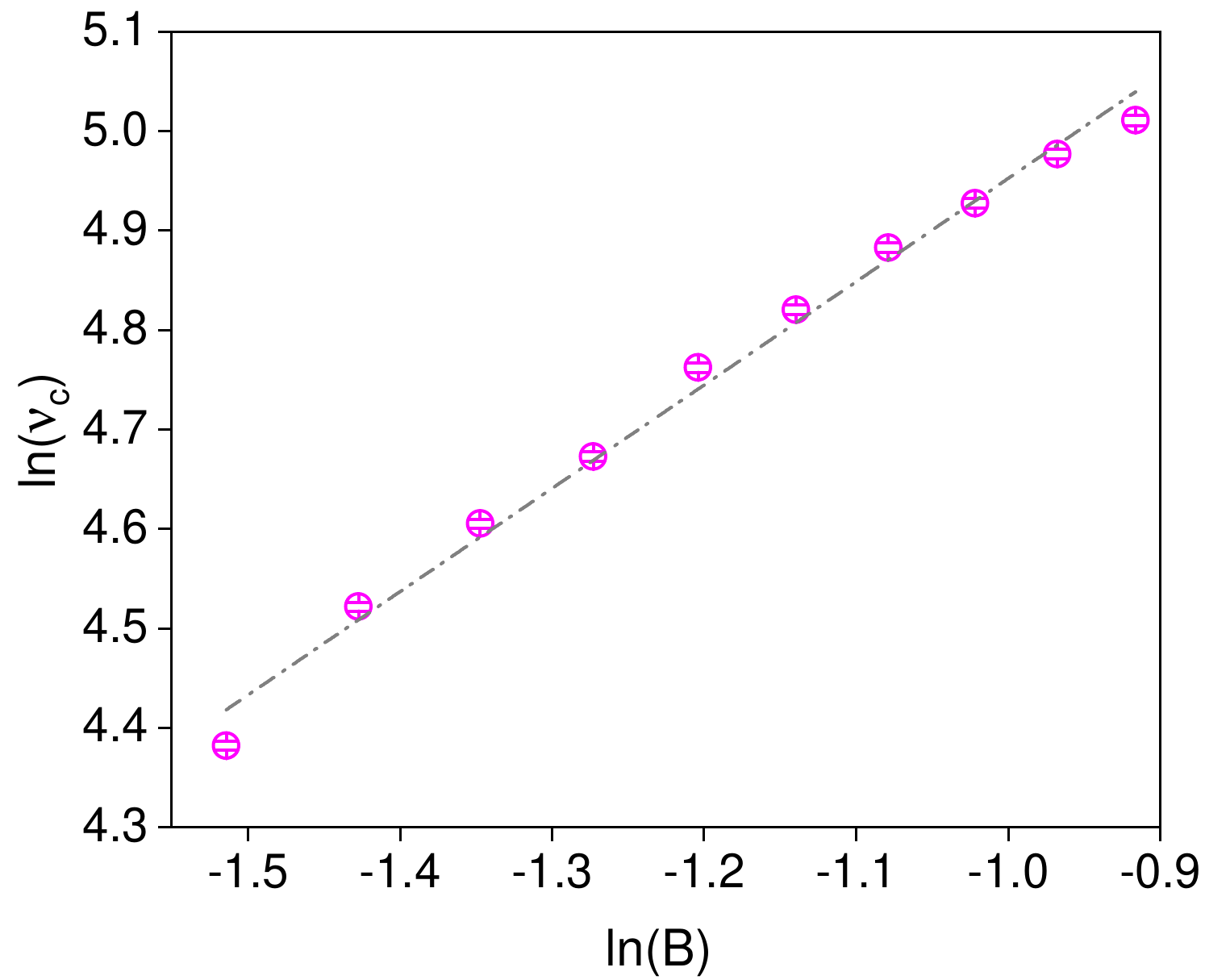}
			\small{\caption{\textbf{$\nu_c$ vs. $B$ plot.} {Plot of filling factor at which node appears $\nu_c$ as a function of $B$ on a ln-ln scale. The black dashed line shows the linear fit to the data with a slope of $\sim 1.03 \pm 0.03$.}  }\label{Fig:S4}}
    \end{center}
\end{figure}

In Fig.~\ref{Fig:S4}, we present the critical filling factor $\nu_c$ versus $B$ plot for device D1. The ${\rm ln}(\nu_c)$ vs ${\rm ln}(B)$ plot shows a linear behavior with a slope of $1.03 \pm 0.03$, consistent with our theoretical analysis.

\section{The effect of in-plane magnetic field on beating}
\begin{figure}
	\begin{center}
			\includegraphics[width=0.5\linewidth]{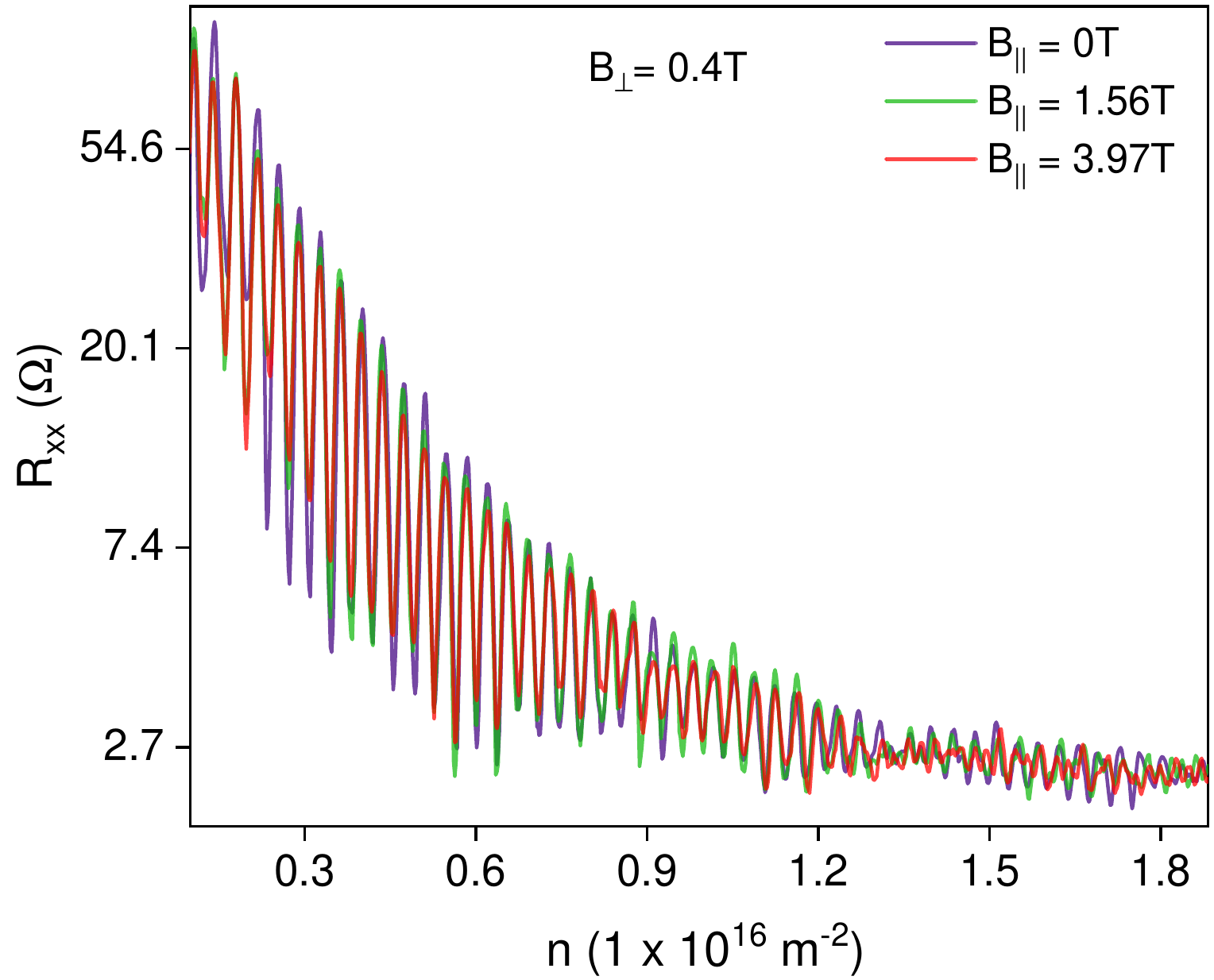}
			\small{\caption{\textbf{Effect of parallel magnetic field on quantum oscillations.} {Plot of the longitudinal resistance as a function of carrier density at a fixed perpendicular magnetic field of $B_{\perp} = 0.4~\mathrm{T}$, measured for various parallel magnetic field strengths. }  }\label{Fig:S3}}
    \end{center}
\end{figure}

In Fig.~\ref{Fig:S3}, we present $R_{xx}(B)$ at a perpendicular magnetic field of $B_\perp = 0.4 $ T, and we vary the applied parallel magnetic field. For these measurements, the device was
rotated with respect to the magnet's axis. We observe from
Fig.~\ref{Fig:S3} shows that for a given value of the perpendicular
component of the magnetic field, the $R_{xx}(B)$ is independent of the parallel component of the magnetic field. This independence validates that the beating does not originate from the spin-split bands.

\section{Estimation of strain and strain gradient from $B_{\mathrm{pm}}$}

\noindent Long-wavelength strain in graphene enters the Dirac Hamiltonian as a valley gauge field:
\begin{equation}
{\bm A}=\frac{\hbar\beta}{2 e a}\,\big(u_{xx}-u_{yy},\,-2u_{xy}\big),
\end{equation}
with lattice constant $a=0.246~\mathrm{nm}$ and Gr\"uneisen parameter $\beta\simeq 2\!-\!3$. The corresponding pseudomagnetic field is:
\begin{equation}
{\bm B}_{\mathrm{pm}}=\nabla\times{\bm A}
=\frac{\hbar\beta}{2 e a}\Big[-\,2\,\partial_x u_{xy}-\partial_y(u_{xx}-u_{yy})\Big].
\end{equation}
For an order-of-magnitude, isotropic estimate, we write
\begin{equation}
B_{\mathrm{pm}}\sim \frac{\hbar\beta}{e a}\,|\nabla\varepsilon|
\quad\Rightarrow\quad
|\nabla\varepsilon|\approx \frac{e a}{\hbar\beta}\,B_{\mathrm{pm}}.
\end{equation}
Here  $|\nabla\varepsilon|$ is the  gradient of the strain components on the scale over which they vary. With $B_{\mathrm{pm}} \approx 13~\mathrm{mT}$,
$|\nabla\varepsilon|
\approx \frac{1.7\times 10^{3}}{\beta/3}\ \mathrm{m^{-1}}$.  Taking $\beta=3$, ${~|\nabla\varepsilon| = (1.7 \pm 0.03)\times 10^{3}\ \mathrm{m^{-1}}~}$. For micron-scale strain variations, the local strain amplitude is of order $10^{-3}$ ($0.1~\%$). For device D2, the $B_{\mathrm{pm}} \sim 1.1~\mathrm{mT}$,
${~|\nabla\varepsilon| = 0.14 \times 10^{3}\ \mathrm{m^{-1}}~}$ giving a local strain of about $0.01 \%$.

\section{Extraction of $B{_{\rm pm}}$ from Fast Fourier transform (FFT)}
For monolayer graphene with (spin and valley) degeneracy, each Landau level (LL) carries four internal degrees of freedom.
At a fixed applied magnetic field $B$, the filling factor is
\begin{equation}
    \nu = \frac{nh}{eB},
\end{equation}
Since each LL changes the filling by $\Delta\nu = 4$, the density period between successive LLs is
\begin{equation}
    \Delta n_0 = \frac{4eB}{h}.
\end{equation}
The corresponding Shubnikov-de Haas (SdH) frequency when oscillations are viewed as a function of density, is therefore
\begin{equation}
    f_0 = \frac{1}{\Delta n_0}
    = \frac{h}{4eB}.
    \label{eq:f0density}
\end{equation}
In the absence of a pseudomagnetic field, the oscillatory part of the longitudinal resistance is approximately periodic in $n$ with frequency $f_0$.

A strain-induced pseudomagnetic field acts with opposite sign in the two graphene valleys.
Denoting the valley index by $\xi=\pm 1$, the effective magnetic fields are
\begin{equation}
    B_\xi = B + \xi B_{\mathrm{pm}},
\end{equation}
Landau quantization then leads to valley-dependent density periods
\begin{equation}
    \Delta n_\xi = \frac{4eB_\xi}{h},
\end{equation}
and hence two distinct SdH density-frequencies
\begin{equation}
    f_\xi
    = \frac{1}{\Delta n_\xi}
    = \frac{h}{4e(B+\xi B_{\mathrm{pm}})} ,
    \label{eq:fpmexact}
\end{equation}
corresponding to the $K$ and $K'$ valley.
Assuming the two valleys contribute with similar amplitude, the oscillatory resistance may be written as a sum of two sinusoidal terms:

\begin{equation}
    \delta R_{xx}(n)
    \propto
    \cos\!\bigl(2\pi f_1 n \bigr)
    +
    \cos\!\bigl(2\pi f_2 n \bigr).
\end{equation}
Using
\[
\cos A + \cos B = 2\cos\!\left(\frac{A-B}{2}\right)
                   \cos\!\left(\frac{A+B}{2}\right),
\]
we obtain
\begin{equation}
    \delta R_{xx}(n)
    \propto
    2\cos\!\Bigl(\pi(f_1 - f_2)\,n\Bigr)\,
      \cos\!\Bigl(\pi(f_1 + f_2)\,n\Bigr).
\end{equation}
The second cosine represents the rapid SdH oscillation, while the first cosine provides the slow envelope (the ``beating'' pattern).
Envelope nodes satisfy
\begin{equation}
    \pi(f_1 - f_2)\,n_j
    = \left(j+\tfrac12\right)\pi,
\end{equation}
implying that the node positions in density are
\begin{equation}
    n_j = \frac{j+\tfrac12}{|f_1 - f_2|}.
\end{equation}
Using Eq.~\eqref{eq:fpmexact}, the difference between the two
density-frequencies is
\begin{align}
    f_1 - f_2
    &= \frac{h}{4e}
       \left(
            \frac{1}{B+B_{\mathrm{pm}}}
            - \frac{1}{B-B_{\mathrm{pm}}}
       \right)
    \\
    &= -\,\frac{h\,B_{\mathrm{pm}}}
              {2e\left(B^2 - B_{\mathrm{pm}}^2\right)}.
\end{align}
Similarly, the sum is
\begin{equation}
    f_1 + f_2
    = \frac{h\,B}
           {2e\left(B^2 - B_{\mathrm{pm}}^2\right)}.
\end{equation}
Taking the ratio yields the exact identity
\begin{equation}
    \frac{f_1 - f_2}
         {f_1 + f_2}
    = -\,\frac{B_{\mathrm{pm}}}{B}.
\end{equation}
Hence, the magnitude of the pseudomagnetic field can be extracted directly
from the splitting of the two FFT peaks:
\begin{equation}
    |B_{\mathrm{pm}}|
    =
    B\,\frac{|f_1 - f_2|}
            {f_1 + f_2}.
    \label{eq:BpmfromFFT}
\end{equation}
Using the above expression, $B_{\rm pm}$ can be directly evaluated from the Fourier transform of the beating pattern.

\section{Intrinsic valley polarization as the origin of beating}
Intrinsic valley polarization refers to an unequal population of carriers in the $K$ and $K'$ valleys, which in graphene layers can arise from broken time-reversal symmetry, such as through magnetic proximity effects, circularly polarized light, or spontaneous valley ordering. This valley imbalance can be characterized by a polarization parameter $p_v \in [-1,1]$. In the low-energy regime, the total carrier density $n$ is split between the two valleys
\begin{equation}
    n_K = \frac{1 + p_v}{2} n, \qquad n_{K'} = \frac{1 - p_v}{2} n.
    \label{eq:valley_densities}
\end{equation}
This gives rise to different oscillation frequencies $f_{0}^K$ and $f_0^{K'}$, since $f_0(E) = E^2/(2e \hbar v_F^2) = \pi n \hbar^2 v_F^2/(2e \hbar v_F^2) = n h/(4e)$. Consequently, the valley-resolved frequencies become
\begin{equation}
    f_0^K = \frac{h (1 + p_v)n}{8e}, \qquad f_0^{K'} = \frac{h (1 - p_v)n}{8e},
\end{equation}
with a frequency difference $\delta f = f_0^K - f_0^{K'} = (h p_v  n)/(4e)$. The interference between these frequencies may, in turn, generate a beating pattern in quantum oscillations.

The total oscillatory signal in DOS and conductivity (see Section~S4) is a sum of contributions from both valleys
\be
    \delta \sigma_{xx}(B) \propto \left[ \cos\left(\frac{2\pi f_0^K}{B}\right) + \cos\left(\frac{2\pi f_0^{K'}}{B}\right) \right] =
     2 \cos\left(\frac{\pi h n}{4e B}\right) \cos\left(\frac{\pi h p_v n}{4e B}\right),
    \label{eq:beating_form}
\ee
where the first term governs the periodic in $1/B$ oscillation and the second term gives a slowly varying amplitude modulation, producing beating.

The beating nodes occur when the slowly varying oscillation term vanishes, i.e.,
\be
\frac{\pi h p_v n}{4e B} = \frac{(2j + 1)}{2} \pi, \qquad j = 0, 1, 2, \cdots.
\ee
This gives the critical carrier density at the nodes to be
\be
n_c = \frac{2 e B}{h p_v} (2j + 1). \label{eq:valley_node_condition}
\ee
This analysis shows that intrinsic valley polarization produces a \textit{linear} dependence of the critical carrier density $n_c$ on magnetic field $B$, in contrast to the \textit{quadratic} scaling $n_c \propto B^2$ observed in our experiment. Therefore, the valley polarization mechanism cannot account for the measured beating behavior.

\section{Ruling out other mechanisms as the source of beating}
A variety of other mechanisms can also generate amplitude modulation or beating-like features in quantum oscillations. For completeness, we review these possibilities, outline their expected signatures, and compare them with our observations. As shown below, none of these mechanisms naturally produces the specific quadratic scaling of the node density or the linear scaling of the filling factor that we measure.

{\textbf{Spin-split bands:}}
Amplitude modulation in oscillatory magnetoresistance can arise from the lifting of spin degeneracy via spin--orbit coupling~\cite{tiwari2023observation,Rao_nc23} or Zeeman splitting in an external field~\cite{hatke2012shubnikov}. In graphene, however, the intrinsic spin--orbit coupling is too small ($\sim \mu$eV) to generate the pronounced beating observed here. Moreover, for spin-split Dirac bands, the condition for beating nodes is mainly independent of carrier density (see Ref.~\cite{Islam_jpcm14_beating}), unlike the precise quadratic scaling of the node position with magnetic field seen in our case. Finally, if the beating originated from spin splitting, it would be susceptible to an in-plane magnetic field. Experimentally, the oscillation and beating pattern remain unchanged under a finite in-plane field (Fig.~S3 of SM~\cite{Note1}), conclusively ruling out both spin--orbit and Zeeman coupling mechanisms.

{\textbf{Multiple Fermi pockets:}}
The periodic modulation of magnetoresistance can also arise from the presence of multiple Fermi pockets of comparable areas at the same energy. Single-layer graphene hosts only a single isotropic Dirac band. This is also evident from the $R_{xx}$ versus $V_{bg}$ plot in Fig.~S1 of SM~\cite{Note1}, where only a single peak of the zero-field resistance is seen around $V_{bg}=0$, corresponding to the Dirac point, ruling out the presence of multiple Fermi pockets in our device.

{\textbf{Intrinsic valley polarization:}}
An intrinsic valley imbalance, arising from spontaneous time-reversal symmetry breaking, lifts the degeneracy between the $K$ and $K'$ valleys by inducing unequal carrier populations. This scenario typically manifests as two inequivalent Dirac points at different energies, resulting in a split-peak structure in zero-field resistivity. However, our $R_{xx}$ versus $V_{bg}$ measurement (Fig.~S1 of SM~\cite{Note1}) displays a single peak centered at charge neutrality, indicating no intrinsic valley imbalance. Moreover, intrinsic valley polarization leads to beating nodes whose critical carrier density scales linearly with magnetic field (see Section~S9 of SM~\cite{Note1}), contrary to the experimentally observed quadratic scaling. Since hBN-encapsulated graphene preserves time-reversal symmetry and lacks any source of spontaneous valley order, this mechanism is incompatible with our system. We therefore rule out intrinsic valley polarization as the origin of the observed beating.

{\textbf{Staggered sublattice potential:}}
Graphene on top of the hBN experiences a staggered potential that can break the A--B sublattice symmetry. The staggered potential $\Delta$ in the Dirac Hamiltonian opens up a uniform band gap $2\Delta$. Firstly, our zero-field $R_{xx}$ measurements versus $V_{bg}$ do not show any appreciable gap in the spectrum. Secondly, the Landau level spectrum in the presence of the gap term, $E_N^{K,K'} = \pm \sqrt{2 e N \hbar v_F^2 B + \Delta^2}$, preserves valley degeneracy. Consequently, the oscillation in the presence of a sublattice potential leads to identical oscillations for both valleys. Hence, a staggered sublattice potential cannot produce a magnetic field-dependent beating~\cite{zeldov_nature23}.

{\textbf{Kekul\'e distortion:}}
The Kekul\'e distortion corresponds to the modulation of hopping amplitude, which can occur in strained graphene~\cite{Eom_nano20} or graphene epitaxially grown on the copper substrate~\cite{Christopher_NP16}. The only effect of O-type distortion is to open up a gap in the band dispersion of graphene. Thus, O-type Kekul\'e distortion can not produce beating in the quantum oscillation. Conversely, the Y-type distortion breaks valley degeneracy~\cite{Gamayun_njp18} by producing different hopping strengths and modifying the Fermi velocity ($v_F \propto$ hopping strength) in two inequivalent valleys. Consequently, the Y-type distortion can generate the beating in oscillation. However, the beating originating from the Y-type Kekul\'e distortion produces a critical filling factor that is independent of the applied magnetic field $B$~\cite{zeldov_nature23}. In contrast, we observe that the critical filling factor depends linearly on $B$, ruling out the Kekul\'e distortion as the possible origin of beating.

{\bf Spatial inhomogeneity or disorder:}
Spatial fluctuations~\cite{Zhang_NP09} in carrier density (e.g., charge puddles) or local strain inhomogeneity can, in principle, result in a distribution of SdH frequencies across the sample, leading to apparent damping or weak interference-like features in the oscillations. However, such effects are not expected to produce the quadratic scaling of the node positions with magnetic field that we observe. In particular, beating due to domains with slightly different carrier densities is predicted to yield node positions that are largely independent of the applied magnetic field~\cite{Taniguchi_acs25}. Moreover, inhomogeneity tends to smear out oscillations rather than generating sharp, well-defined, and periodic beating nodes. In contrast, all our devices exhibit highly regular and reproducible beating patterns. Finally, the high carrier mobility ($3.5 \times 10^5~\rm cm^2V^{-1} s^{-1}$) in our devices implies long quantum lifetimes and minimal disorder, further ruling out inhomogeneity as the source of the observed beating.

{\textbf{Magnetic breakdown:}}
Magnetic breakdown refers to quantum tunneling between adjacent Fermi orbits in momentum space when the cyclotron energy becomes comparable to their energy separation \cite{Shoenberg1984}. Such tunneling can induce interference between quantized orbits, leading to amplitude modulation in quantum oscillations. In single-layer graphene, however, each valley hosts a single Fermi pocket, and the valleys are separated by a large momentum, $\delta k = \tfrac{4\pi}{3a} \approx 1.7$ \AA$^{-1}$ (with $a$ being the lattice constant). Since the breakdown field scales as $B_{\mathrm{MB}} \sim  (h/e)\, \delta k^2$~\cite{Shoenberg1984, Zeldov_sc24}, the characteristic field is of the order of $10^6$ T, far beyond the experimentally accessible range. Consequently, no evidence of magnetic breakdown has been observed in single-layer graphene, effectively ruling out this mechanism.

These analyses rule out all mechanisms mentioned above as potential explanations for the observed beating in the SdH oscillation in our devices. Instead, the only consistent interpretation of our data is the emergence of valley-dependent Landau quantization due to strain-induced pseudomagnetic fields in the presence of an external magnetic field.

\bibliographystyle{apsrev4-2}
\bibliography{arxiv}

\end{document}